\documentclass[sigconf]{acmart}
\setcopyright{none}

\newcommand{\fundwo}[1]{#1}

\AtBeginDocument{%
  }

\usepackage{booktabs}
\usepackage{multirow}
\usepackage{tabularx}
\usepackage{array}
\usepackage{amsmath}
\usepackage{graphicx}
\usepackage{makecell}
\usepackage{placeins} 
\begin{document}

\title{MIDI-Informed Singing Accompaniment Generation in a Compositional Song Pipeline}

\author{Fang-Duo Tsai}
\affiliation{%
  \institution{National Taiwan University}
  \institution{Taiwan AI Labs}
  \city{Taipei}
  \country{Taiwan}
}
\email{fundwotsai2001@gmail.com}

\author{Yi-An Lai}
\affiliation{%
  \institution{National Taiwan University}
  \city{Taipei}
  \country{Taiwan}
}

\author{Fei-Yueh Chen}
\affiliation{%
  \institution{University of Rochester}
  \city{Rochester}
  \state{New York}
  \country{USA}
}

\author{Hsueh-Wei Fu}
\affiliation{%
  \institution{National Taiwan University}
  \city{Taipei}
  \country{Taiwan}
}
\author{Wei-Jaw Lee
}
\affiliation{%
  \institution{National Taiwan University}
  \institution{Taiwan AI Labs}
  \city{Taipei}
  \country{Taiwan}
}
\author{Hao-Chung Cheng}
\affiliation{%
  \institution{National Taiwan University}
  \city{Taipei}
  \country{Taiwan}
}

\author{Yi-Hsuan Yang}
\affiliation{%
  \institution{National Taiwan University}
  \institution{Taiwan AI Labs}
  \city{Taipei}
  \country{Taiwan}
}

\renewcommand{\shortauthors}{Tsai et al.}

\begin{abstract}
While end-to-end lyrics-to-song models offer convenience for casual users, professional songwriters require score-to-song systems that allow them to retain authorship over the core melody. However, existing score-to-song methods are limited to short-form snippets and fail to maintain coherence in long-form generation, particularly during vocal-silent sections like intros and bridges.
To address this long-form bottleneck, we propose MIDI-informed singing accompaniment generation (MIDI-SAG). Unlike conventional audio-only models, MIDI-SAG utilizes symbolic timing and chord information derived from the vocal MIDI to provide a stable musical roadmap. By incorporating structure planning, which defines temporal boundaries and semantic labels, our framework facilitates consistent generation across both vocal and non-vocal sections. We demonstrate the feasibility of this compositional pipeline by leveraging specialized pre-trained modules, enabling data-efficient training on a single GPU. Our experiments show the potential of this approach for both professional score-to-song and general lyrics-to-song tasks. While an early exploration, MIDI-SAG suggests a promising direction for structured, long-form music synthesis. Audio demos are available, and the code will be open-sourced at \url{https://composerflow.github.io/web_revealed/}.
\end{abstract}





\maketitle

\section{Introduction}

Recent advances in generative music models have led to commercial systems (e.g., \citet{suno_v4_5_2025}) that demonstrate high-quality, convenient generation of musical audio pieces that involve both the singing voices and instrumental accompaniment. In general, we refer to such systems as \emph{song generation} systems. Depending on the input of such systems, however, we can further distinguish between two types of song generation systems, as described below. 

The first type, referred to as \emph{lyrics-to-song}  systems here, assumes that the input is a nature language description of the music (i.e., a text prompt) and the intended lyrics, and the system would handle the composition, arrangement, and performance simultaneously.
This is the scenario assumed by Suno and emerging open-source variants such as Yue~\cite{yuan2025yue}, DiffRhythm~\cite{ning2025diffrhythm}, and ACE-Step~\cite{gong2025ace}.
These systems typically adopt a monolithic, end-to-end approach, mapping lyrics and text descriptions directly to audio. 
Because the input is simple, such systems can be easily used by casual users and content creators without profound musical knowledge. 

\begin{figure}[t]
    \centering
    \includegraphics[width=\columnwidth]{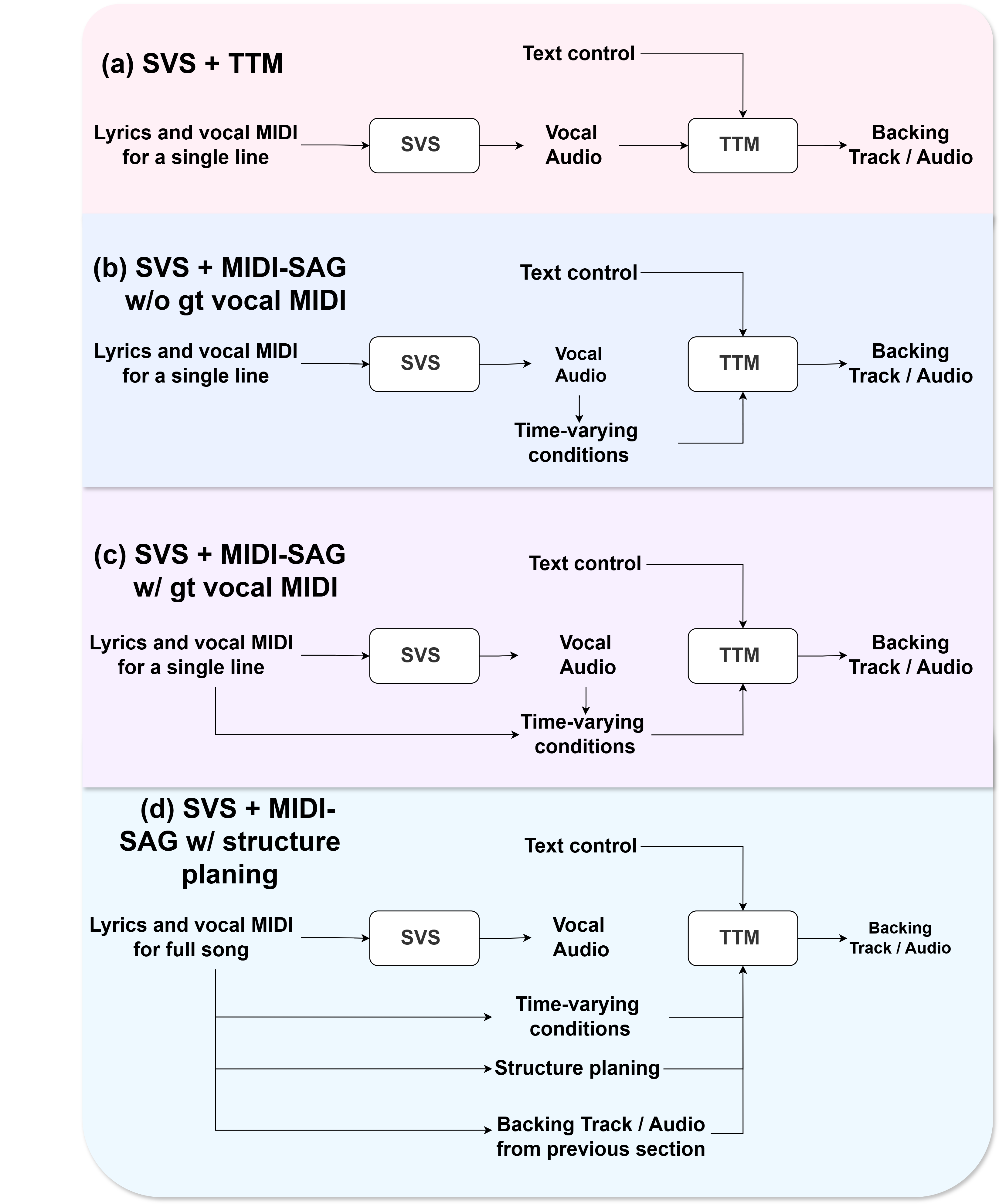}
    \caption{Architectural comparisons of score-to-song generation systems: (a) na\"ive chaining of SVS and SAG as done in \cite{melodist24acl}, which adapts a text-to-music (TTM) model for SAG; (b) the proposed extension (cf. Section~\ref{SAG}) that uses score-related time-varying conditions extracted from the vocal audio for SAG (named MIDI-SAG); (c) the proposed extension  (cf.~\ref{SAG},~\ref{harmonization}) that gets the conditions instead from the input MIDI score; (d) the proposed extension  (cf.~\ref{completeness}) that considers full-song MIDI and structure planning.}
    \label{SAG_variants}
\end{figure}

In contrast, the second type of systems, \emph{score-to-song} systems, require a user to provide additionally the MIDI score of the singing voice, specifying the pitch, duration of each melody note. As such, the target users are professionals who desire to retain authorship over the core singing melody. Melodist~\cite{melodist24acl} represents a notable example of such a system. It combines a singing voice synthesis (SVS) module~\citep{lu2020xiaoicesing,chen2020hifisinger,liu2022diffsinger} with a singing accompaniment generation (SAG) module~\citep{donahue2023singsong, chen2024fastsag, trinh2024sing, tsaidemonstrating} , using the SVS to firstly generate the vocal track (i.e., the singing voice) according to the musical score and  lyrics, and then the SAG to generate the backing track (i.e., the instrumental accompaniment) given the text prompt and vocal track, as conceptually depicted in Figure~\ref{SAG_variants}(a). The summation of the two tracks is the final output of the system.

While sharing the same goal of producing songs, the two types of systems clearly differ in their level of creative agency and the technical entry barrier for the user. 
Possibly due to accessibility and market reasons, however,
the progress of score-to-song generation has greatly lagged behind that of lyrics-to-song generation.
While the idea of combining SVS and SAG to realize score-to-song generation is straightforward and easy to come by, Melodist~\cite{melodist24acl} represents the only attempt in the recent literature, to our best knowledge. Moreover, as Melodist follows the conventional SVS approach and generates the singing \emph{line-by-line} (i.e., one sentence at a time), it can only generate 10-second \emph{short-form} audio snippets, not \emph{long-form} music (e.g., up to two minutes) as offered by state-of-the-art lyrics-to-song generation systems.


The goal of this paper is to advance score-to-song generation by addressing this \emph{long-form  bottleneck}. 
Specifically, we argue that long-form score-to-song generation cannot be achieved simply by chaining SVS and SAG as done in the prior work ~\cite{melodist24acl}. 
The rationale is simple: SAG models often assume that the vocal is present in the input, but this is not the case for long-form music as there are vocal-silent (non-vocal) sections such as intro, bridge, and outro. 
While lyrics-to-song systems handle these sections implicitly, score-to-song systems relying on the conventional SAG approach fail when the primary conditioning signal (i.e., the vocal) is absent.  Therefore, simply extending the SVS input to multiple lyric lines cannot work well.
The SAG part and the whole compositional pipeline itself have to be re-designed for long-range coherence.

We propose three approaches to improve score-to-song generation. 
First, as shown in Figure~\ref{SAG_variants}(b),  while conventional SAG approaches (named \textbf{audio-SAG} hereafter) \citep{donahue2023singsong, chen2024fastsag, trinh2024sing, zhang2025versatile, li2024accompanied, melodist24acl} use the  vocal audio as the input,
we propose to use instead time-varying conditions such as \emph{melody} and \emph{beats} extracted from the vocal audio as the SAG input. 
Compared to the raw audio, such ``content-related'' time-varying conditions from the vocal track may provide more explicit conditions for the generation of the accompaniment track. 

Second, as shown in Figure~\ref{SAG_variants}(c),  we later recognize that, in the context of score-to-song generation, most content-related time-varying conditions are actually \emph{readily} available from the MIDI score, which is part of the input of the SVS module (but conventionally \emph{not} an input of the SAG module). Accordingly, instead of estimating these content conditions from the vocal audio, which is not error-free, we can compute the content conditions directly from the MIDI score. 
This approach significantly bolsters \emph{rhythmic} consistency, as we can more easily trace the beats and downbeats from the MIDI  than from the vocal audio~\cite{heydari2022singing}. Moreover, we enhance \emph{harmonic} consistency between the vocal and backing tracks by integrating a \emph{melody harmonization} module~\cite{yi2022accomontage2} to derive chord progressions from the MIDI. This dual-conditioning—symbolic beats and chords—provides SAG with a structural roadmap unavailable in traditional audio-only approach.
We refer to this ``cross-modality, MIDI-to-audio'' generation process as \emph{MIDI-informed} singing accompaniment generation, or \textbf{MIDI-SAG} for short. We refer to what depicted in Figure~\ref{SAG_variants}(c) as the default case of MIDI-SAG ``with groundtruth vocal MIDI,''
and the one in (b) as the alternative case ``without groundtruth vocal MIDI.'' 
Please note that, in either (b) or (c), the vocal audio itself is not taken directly as the input to the proposed (MIDI-)SAG module.

Finally, with MIDI-SAG in place, we propose the approach depicted in Figure~\ref{SAG_variants}(d), which explicitly tackles the long-form bottleneck.
Specifically, from the full-song lyrics and MIDI,  we derive a structural planning specifying the global musical form. Each section is defined by its temporal boundaries and semantic labels (e.g., intro, verse). 
Moreover, we derive the beat positions and chord progression for the whole song, for both vocal and non-vocal sections. We perform SAG with the section boundaries in mind, generating the backings for the vocal sections first, and then for the non-vocal sections. 
This way, even the vocal part is silent in non-vocal sections, we can leverage the global rhythmic and chordal conditions, along with the audio outpainting capabilities of TTM models~\cite{tsai2025musecontrollite}, to realize coherent long-form generation. 

We demonstrate the feasibility of this compositional pipeline by leveraging a pre-trained SVS module~\cite{ren2020fastspeech} and a pre-trained TTM module (for SAG)~\cite{evans2025stable}, finetuning only the SAG part by adding lightweight adapters~\citep{wu2024music,tsai2025musecontrollite} to incorporate various time-varying conditions.
In our implementation, the finetuning process only requires 2.5k hours of audio data and a single RTX 3090 GPU.
We report experiments comparing the architectures shown in Figure~\ref{SAG_variants} for score-to-song generation, validating the effectiveness of the proposed approaches.


Interestingly, besides score-to-song generation, our compositional pipeline can be extended to support lyrics-to-song generation. We consider this as a ``bonus'' task and implement it by adding a lyrics-to-melody module CSL-L2M~\cite{chai2025csl} to our pipeline, so that the vocal MIDI can be \emph{generated} rather than \emph{provided} at the input side of the overall system.
We also perform experiments comparing this prototype system with state-of-the-art lyrics-to-song generation systems, and we see promising initial result.

Our contributions are three-fold: (i) MIDI-SAG, a new way to feed symbolic MIDI priors into generative audio models to solve the long-form bottleneck; (ii) structural completeness, supporting intermittent vocals for consistent backing in long-form generation; and (iii) a data-efficient compositional pipeline that exhibits promising results for both score-to-song and lyrics-to-song generation.


\section{Related Work}
Musical audio generation has evolved rapidly, with \emph{text-to-music} (TTM) models \cite{agostinelli2023musiclm, musicgen, liu2024audioldm, fei2024flux,novack25presto,yang25tvsmusicgen,zhang25inspiremusic,evans2025stable,niu26steermusic, tsai2024audio, lee2026training} focusing primarily on generating instrumental music.
In contrast, \emph{lyrics-to-song} generation produces integrated vocal and instrumental audio. Following early work like Jukebox \citep{dhariwal2020jukebox}, and commercial systems like \citet{suno_v4_5_2025}, a proliferation of open-source models has emerged, including autoregressive frameworks~\citep{yuan2025yue,lei2025levo,yang2026heartmula}, and diffusion-based architectures~\citep{ning2025diffrhythm, jiang2025diffrhythm, gong2025ace,yang2025songbloom}.
Despite their high fidelity, these end-to-end approaches suffer from prohibitive computational costs, monolithic architectures, limited user control, and the lack of transparency. 

Compared to either TTM or lyrics-to-song generation, the \emph{score-to-song} generation task receives relatively less attention, as the target users are more for professionals or MIDI-capable hobbyists who want to hear their specific MIDI composition realized.
Melodist~\cite{melodist24acl} pioneered building such a system in early 2024, but most (if not all) of the subsequent song generation models are for lyrics-to-song generation \cite{yuan2025yue,ning2025diffrhythm,gong2025ace,lei2025levo,liu2025songgen, yang2025songbloom,lei2025levo, liu2025jam, yang2026heartmula}, not supporting score input. 
We note that, while most of them adopt an \emph{end-to-end} approach, 
MelodyLM~\cite{li2024accompanied}  and VersBand~\cite{zhang2025versatile} use a modular approach integraging lyrics-to-melody, SVS, and SAG modules for \emph{compositional} lyrics-to-song generation.
Due to their modular design, we can simply drop the lyrics-to-melody module and use user-provided melody score as the system input to realize score-to-song generation (e.g., though this is not their original target). 
However, either Melodist, MelodyLM or VersBand, generates snippets of only 20 seconds or shorter, and ignores entirely the challenges associated with long-form generation.




Lastly, focusing on SAG alone, we note that existing SAG models~\citep{donahue2023singsong, chen2024fastsag, trinh2024sing, zhang2025anyaccomp, zhang2025versatile} assume exclusively vocal audio as the input and deal with an \emph{audio-to-audio} generation task.
This is in stark contrast with the proposed MIDI-SAG formulation, which deals with a novel \emph{MIDI-to-audio} generation task leveraging symbolic MIDI input either estimated from the vocal audio or readily available from preceding SVS module in a compositional pipeline.

\section{Methodology}

\subsection{System Overview \& Professional Workflow}

To address the limitations of existing end-to-end models, we propose a compositional song pipeline specifically designed for a score-to-song professional workflow. 
Specifically, a user of our system would provide a full-song lyrics $L$ providing  the song form (section labels) and 
the lyrics line-by-line, the corresponding vocal MIDI score $M$ specifying the pitch and duration of each melody note and their association with each word in the lyrics, as well as a text description $D$ describing the intended overall style, mood or instrumentation of the song. 
Such an input tuple $\{L,M,D\}$ ensures that the professional songwriter retains complete authorship over the melodic content and rhythmic phrasing of the composition.

As illustrated in Figure~\ref{SAG_variants}(d), our system decomposes the complex task of long-form song generation into a sequence of specialized, controllable modules. The workflow follows a five-stage process:
\begin{enumerate}
  \item Structure planning: Given the input $L$ and $M$, the system establishes a global structure plan ($S$), which defines the temporal boundaries (start and end timestamps) and semantic labels (e.g., intro, verse) for the entire song.
  \item Vocal synthesis: A pre-trained SVS module converts the lyrics and MIDI into a high-fidelity vocal audio track ($V$).\item Harmonic roadmapping: An off-the-shelf harmonization module analyzes the melody $M$ to suggest a compatible chord progression ($C$) for the whole song.
  \item MIDI-informed generation: The core MIDI-SAG module synthesizes the accompaniment audio ($A$). Unlike traditional audio-only SAG, our MIDI-SAG is an adapted TTM that is conditioned on $D$, $M$, $C$, and $S$ simultaneously.
  \item Long-form inference: Using an audio continuation/outpainting strategy, the system generates the final mixed song in a chunk-wise fashion while maintaining global coherence via the structural anchors established in the previous stages.
\end{enumerate}

A key advantage of this modular approach is its fine-grained editability. At any stage before the final synthesis, a professional user can manually adjust the lyrics, MIDI score, structure plan, the synthesized voices, or the suggested chord progression. Furthermore, by leveraging specialized pre-trained modules, the primary generative component (i.e., MIDI-SAG) can be trained efficiently using lightweight adapters on a single consumer-grade GPU.

In the following subsections, we provide the technical details of our three core extensions: MIDI-informed conditioning (Section~\ref{SAG}), explicit harmonic guidance (Section~\ref{harmonization}), structural planning and long-form inference for long-range coherence (Section~\ref{completeness}).

\subsection{MIDI-informed SAG}\label{SAG}

Given the synthesized vocal $V$, the SAG module generates the complementary accompaniment $A$ to produce the final mix. 
This task requires precise temporal synchronization of beat and downbeat structures, alongside harmonic alignment between the vocal melody and the underlying instrumental progression.
Existing audio-based SAG models~\cite{donahue2023singsong, chen2024fastsag, trinh2024sing, zhang2025versatile, li2024accompanied}, including the na\"ive adaptation of TTM for SAG attempted in the prior work Melodist~\cite{melodist24acl} (see Figure~\ref{SAG_variants}(a))
attempt to solve this audio-to-audio generation task in an end-to-end fashion.
However, this approach necessitates massive datasets—often exceeding 40k hours~\citep{donahue2023singsong}—to learn the mapping.
Our preliminary studies indicate that under resource-constrained conditions (e.g., 2.5k hours of data and a single GPU), audio-only conditioning fails to achieve sufficient coherence.

In contrast, the proposed MIDI-SAG utilizes symbolic priors from the content conditions to simplify the learning task. 
Specifically, such content-related, time-varying conditions may include the melody line and beat positions of the singing voice, with which the SAG may find easier to generate a backing track that is rhythmically and harmonically  cosistent with the vocal track.
While early TTM or SAG do not consider such time-varying conditions, recent work on controllable TTM, such as Music ControlNet~\citep{wu2024music} and MuseControlLite~\citep{tsai2025musecontrollite}, has established the methodology to incorporate time-varying controls to a music generation model via lightweight adapters, paving the way to develop our MIDI-SAG.


We consider two variants of MIDI-SAG in this paper. 
The first approach assumes that the vocal MIDI $M$ is not available and infer the conditions directly from the vocal audio $V$, as depicted in Figure~\ref{SAG_variants}(b). 
In our implementation, we employ the off-the-shelf MIDI extractor SOME~\cite{openvpi_some} to extract the symbolic melody (i.e., not the F0 contour),
and a custom-trained singing beat detector following~\citet{heydari2022singing} for  beat detection.\footnote{To ensure stability across the entire song, we address a key limitation of beat detectors: their tendency to produce random activations during vocal silence. Specifically, we integrate Silero VAD~\citep{silero_vad} to distinguish between vocal and non-vocal segments, and interpolate beat timestamps for the non-vocal parts using the tempo value inferred from neighboring vocal regions.}

The second MIDI-SAG variant, which we consider as the default case, leverages the compositional pipeline’s inherent data coupling and compute the conditions directly from the vocal MIDI score $M$, as depicted in Figure~\ref{SAG_variants}(c). 
This strategy is advantageous in that it ensures \emph{rhythmic precision} by utilizing the unambiguous beat and downbeat timings inherent in the symbolic melody, bypassing the noise and estimation errors typical of raw audio analysis.
Moreover, it allows for creating explicit and accurate harmonic guidance for SAG, as described in the next subsection.

\begin{figure}[t]
    \centering
    \includegraphics[width=.75\columnwidth]{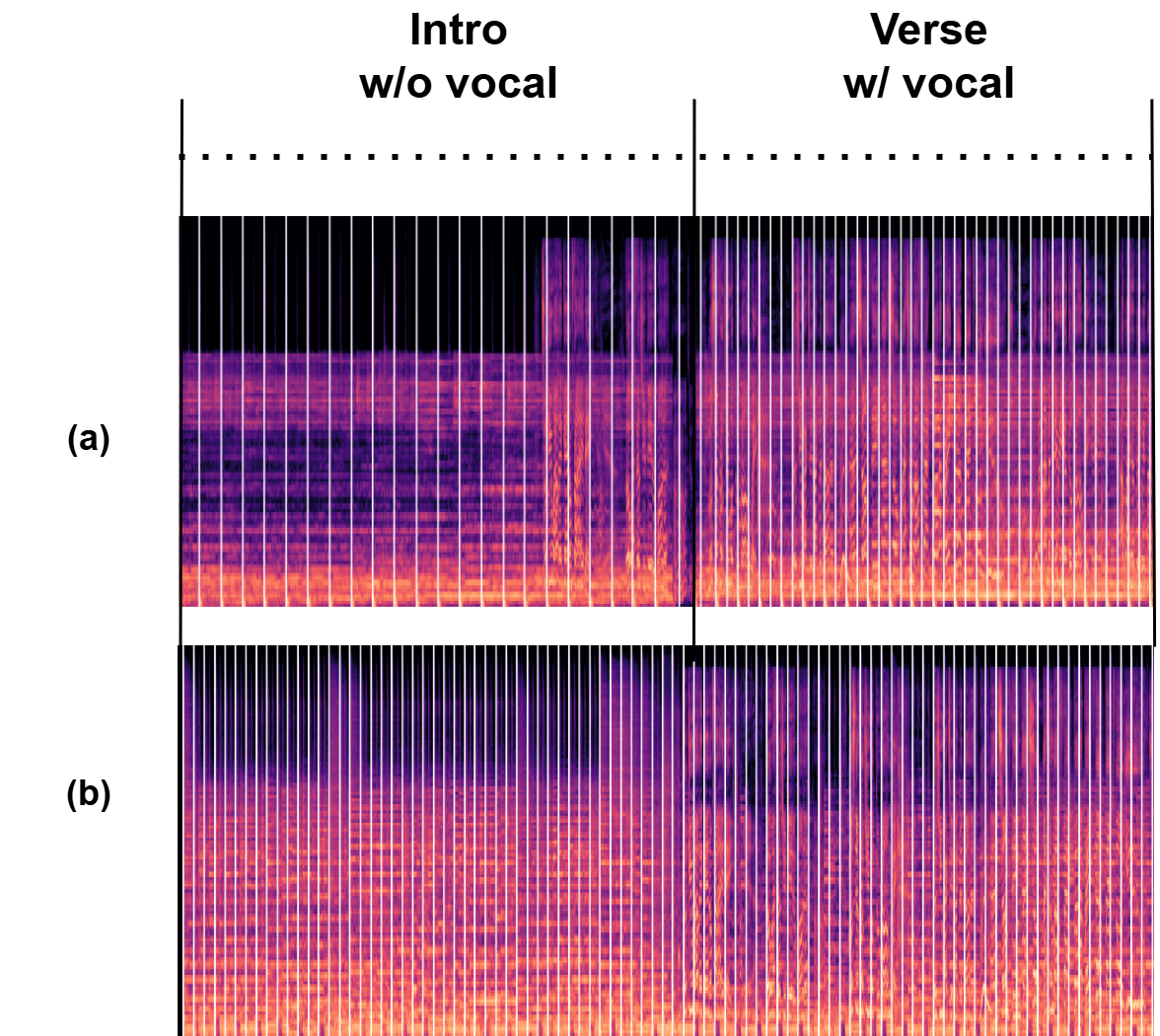}
    \caption{Illustration of rhythmic stability. White stripes represent the predicted beat positions from a generated accompaniment. While (a) audio-SAG loses rhythmic consistency between the non-vocal (first-half), vocal (second-half) segments, (b) MIDI-SAG yields stable beat and coherent content.}
    \label{fig:sag-comparison}
\end{figure}

\subsection{Explicit Harmonic Guidance}\label{harmonization}

We consider two types of time-varying conditions in Section \ref{SAG}: 
\emph{melody} and \emph{beats}.
Given the beat conditions from the groundtruth MIDI, ideally our MIDI-SAG  can simply follow the beat conditions to generate a backing track that is rhythmically consistent with the vocal track, since the two tracks are  temporally synchronized.

However, the melody condition alone is insufficient to guarantee that the generated backing is harmonically coherent with the vocal.
In music theory, the relationship between a melody and its accompaniment is not a simple one-to-one mapping~\cite{chuan07,temperley_2007}. A single melodic line can be harmonized in numerous ways.
Without explicit harmonic guidance, the generated backing may fail to provide the necessary functional harmonic support for the vocal.

To address this, we propose to additionally a time-varying \emph{chord} condition for the SAG. 
While we cannot directly estimate the chord condition from either the monophonic vocal audio or MIDI score, we can use off-the-shelf melody harmonization module~\cite{yi2022accomontage2}
to suggest a chord progression $C$ given the vocal melody line, improving the harmonic consistency of the generated backing.
Moreover, making the chord condition explicit is advantageous in our professional workflow, since the creator may have a specific tonal center and chordal structure in mind to be enforced to the song.

We note, while the melody condition is only available for vocal sections, both the beat and chord conditions extend well to the non-vocal sections.
Conventional audio-SAG models, lacking the symbolic guidance, often struggle to maintain coherence during vocal silences. In contrast, our MIDI-SAG generates 
musically consistent result. See Figure~\ref{fig:sag-comparison} for an exemplar  instrumental intro.

\begin{figure*}[t]
    \centering
    \includegraphics[width=1\textwidth]{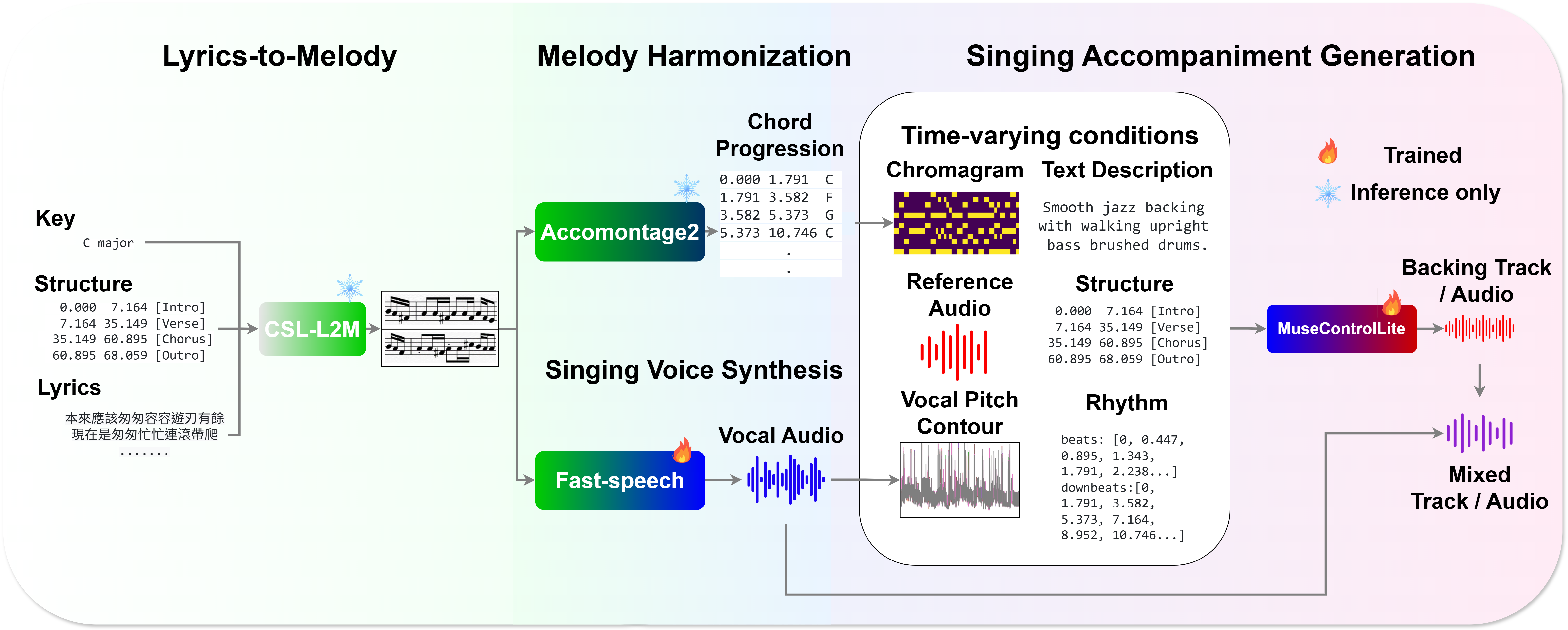} 
    \caption[c]{Overview of the implemented compositional song generation pipeline. It performs \emph{score-to-song} generation when the melody score is provided by a user (i.e., without the lyrics-to-melody module), and performs \emph{lyrics-to-song} generation when the melody is generated by the lyrics-to-melody module. 
    For lyrics-to-song, the system sequentially maps lyrics to a full song through: (1) melody composition (CSL-L2M~\citep{chai2025csl}), (2) SVS (FastSpeech~\cite{ren2020fastspeech}), (3) melody harmonization (AccoMontage2~\citep{yi2022accomontage2}), and (4) the proposed MIDI-SAG, which adapts MuseControlLite~\cite{tsai2025musecontrollite} to incorporate symbolic and acoustic conditions.}
    \label{ComposerFlow}
\end{figure*}

\subsection{Structural Planning \& Long-Form Inference}\label{completeness}

Most generative models, including modern Transformer-based or diffusion-based architectures, are subject to a fixed maximum sequence length. For instance, Stable Audio Open\,(SAO)~\citep{evans2025stable} has a receptive window of 47 seconds. Consequently, to generate a multi-minute music, we often need multiple inference passes. 
Maintaining structural integrity over extended durations is therefore a challenge for long-form generation. Moreover, conventional SAG models are typically trained on short audio clips and assume the continuous presence of a vocal signal. These models may fail when tasked with generating songs that include instrumental, non-vocal sections.

To overcome this challenge, we employ a multi-pass inference strategy guided by a structure plan $S$, as illustrated in Figure~\ref{SAG_variants}(d).  
Instead of attempting to synthesize the entire track at once, our MIDI-SAG generates the accompaniment in finite-duration, overlapping windows. 
To ensure seamless transitions between these windows, we utilize an audio outpainting technique where the final (or first) seconds of the previous (subsequent) segment serve as an acoustic prefix (postfix) for the next pass.
For each window, the model is provided with the time-varying conditions (e.g., beats and chord) corresponding to the specific segment to serve as the symbolic roadmap for generation.
Moreover, we propose the following two techniques for better long-range coherence.

\emph{Section-anchored slicing}:  
Rather than slicing  at arbitrary time points, we segment a song into multiple windows according to the structure plan $S$, which is a sequence of  segments, each defined by a section label $l_i$ (from $L$) and its corresponding temporal boundaries $[t^{start}_i, t^{end}_i]$ (from $M$). These labels categorize sections of the song into functional units such as intro, verse.
By anchoring a window at the start of a functional section, the model learns to synthesize transitions that respect the song's architectural layout, using the preceding section's audio as a reference for the current generation.


\emph{Backward outpainting} addresses the ``cold start'' problem during instrumental intros, where the absence of vocal pitch signals leads to impoverished conditioning. During training, if a target window begins with an intro, we replace the reference audio with the subsequent vocal section with a 50\% probability. This strategy encourages the model to maintain stylistic and harmonic consistency across the entire track, effectively allowing the tonal characteristics of the main song body to inform the preceding instrumental sections.

\subsection{Extension to Lyrics-to-Song Generation}\label{lyrics2song}

Figure~\ref{ComposerFlow} provides an overview of the proposed song generation system. 
With SVS, melody harmonization, and the core MIDI-SAG modules, our system performs score-to-song generation, generating the vocal and backing tracks $V$ and $A$ given $\{L,M,D\}$.

Interestingly, we note that, by adding a lyrics-to-melody module that generates $M$ given $L$, our system can be extended to realize a compositional lyrics-to-song generation pipeline, as also depicted in Figure~\ref{ComposerFlow}. In this case, the system generates $V$ and $A$ given $\{L,D\}$.

\begin{table}[t]
  \centering
  \caption{Conditioning signals used during training and inference times for the proposed MIDI-SAG module.}
  \label{tab:conditioning-signals}
  \small
  \setlength{\tabcolsep}{4pt}
  \renewcommand{\arraystretch}{1.05}
  \fundwo{
  \begin{tabular}{@{}l p{0.29\linewidth} p{0.36\linewidth}@{}}
    \toprule
    \textbf{Condition type} & \textbf{Training } & \textbf{Inference } \\
    \midrule
    beat &
      All-in-One~\citep{kim2023all} &
      From generated vocal MIDI \\
    pitch &
      RMVPE~\citep{wei2023rmvpe} &
      RMVPE~\citep{wei2023rmvpe} \\
    chord &
      Chord detection~\citep{ParkCJKP19} &
      Accomontage2~\citep{yi2022accomontage2} \\
    structure &
      All-in-One~\citep{kim2023all} &
      User-provided or by LLM \\
      \midrule
    text prompt &
      Audio\,Flamingo\,3~\cite{goel2025audio}  &
      User-provided or by LLM \\
      \midrule
    reference audio &
      Mask one structure; others as reference &
      From previous song-structure segment \\
    \bottomrule
  \end{tabular}
  }
\end{table}

\section{Implementation Details}\label{implement}

As shown in Figure~\ref{ComposerFlow}, we used FastSpeech~\cite{ren2020fastspeech} for SVS, Accomontage2~\cite{yi2022accomontage2} for melody harmonization, and the proposed MIDI-SAG for SAG. We optionally
employed CSL-L2M~\citep{chai2025csl} for lyrics-to-melody for the experiment on lyrics-to-song generation.

We trained FastSpeech~\citep{ren2020fastspeech} from scratch on a 10-hour internal corpus from two licensed singers (male/female), with aligned MIDI, phonemes, and mel spectrograms. The model was trained for 24 hours on a single NVIDIA RTX 3090.


We used the pre-trained latent diffusion model SAO~\citep{evans2025stable} for our SAG, and 
the MuseControlLite~\citep{tsai2025musecontrollite} adapters to inject time-varying controls, which are temporally interpolated and concatenated along the cross-attention feature dimension. 
The original MuseControlLite uses separate adapters for the reference audio and time-varying control, arguing that  there would be abrupt changes at the transition boundary otherwise. We found that, however, we can use a single set of adapters to smoothly model all the conditions, with the following modifications: (i) by unfreezing the self-attention layers in the pretrained backbone, and (ii) by performing audio outpainting at structure boundaries (from our structure plan $S$) rather than at arbitrary timestamps. Under this design, the multiple classifier-free guidance formulation can be simplified as follows:
\begin{equation}\label{eqn:multi-cfg}
\begin{aligned}
&\nabla_{x} \log p_{\lambda}(\mathbf{x} | \mathbf{c}) = \nabla_{x} \log p(\mathbf{x}) \; + \\
&\quad \sum_{i \in \{\text{text}, \text{time-varying}\}} \lambda_i \Bigl(\nabla_{x} \log p(\mathbf{x} | \mathbf{c}_{\leq i}) - \nabla_{x} \log p(\mathbf{x} | \mathbf{c}_{< i})\Bigr).
\end{aligned}
\end{equation}
We fine-tuned the SAO on 2.5k hours of Mandarin pop using the same RTX 3090 GPU for 9 days. 
Additionally, to support creative transitions among different sections using different text prompts, we modified the original cross-attention mechanism of the SAO backbone, and trained it with the MuseControlLite~\cite{tsai2025musecontrollite} adapters:
\begin{align}
\text{Attn}(\mathbf{Q},\mathbf{K},\mathbf{V})
= \text{softmax}~\!\left(\frac{\mathbf{Q}\mathbf{K}^\top}{\sqrt{d}}\right)\mathbf{V} \,, \\
\mathbf{H}
= \text{Concat}~\!\Big(
\text{Attn}(\mathbf{Q}_{:m}, \mathbf{K}_1, \mathbf{V}_1),\,
\text{Attn}(\mathbf{Q}_{m:}, \mathbf{K}_2, \mathbf{V}_2)
\Big)
\end{align}
where $\mathbf{K}_1, \mathbf{V}_1$, $\mathbf{K}_2, \mathbf{V}_2$ are the keys and values from the text prompt of the current and next sections, respecitvely, and $m$ denotes the boundary of the transition. 
Specifically, we performed structure analysis using All-in-One~\cite{kim2023all} to get the structure plan. Since the instrumentation often changes during transitions, we then used Audio\,Flamingo\,3~\cite{goel2025audio} to derive captions for each song section. This supports the above multi-text cross-attention control, since the 47-second window of SAO often includes multiple song sections.


\begin{table*}[t]
\centering
\caption{Result of Experiment 1, comparing different SAG settings for score-to-song generation, using the same SVS module. The scores are the higher the better for all metrics. Coherence, Musicality, Memorability, Clarity, Naturalness are evaluated using SongEval~\cite{yao2025songevalbenchmarkdatasetsong}. The vocal audio starts with two silent bars as the intro.}
\label{tab:midi_sag_results}
\setlength{\tabcolsep}{5pt}
\resizebox{\textwidth}{!}{
\begin{tabular}{lrccc|ccccc}
\toprule
\textbf{SAG Method}& \textbf{duration} & \textbf{Rhythm F1} & \textbf{Key Acc.} & \textbf{CLAP} & \textbf{Coherence} & \textbf{Musicality} & \textbf{Memorability} & \textbf{Clarity} & \textbf{Naturalness} \\
\midrule
TTM with no vocal input                              & 47s   & 0.2061 & 0.0650 & \textbf{0.3221} & 2.9056 & 2.8372 & 2.6813 & 2.7236 & 2.6762 \\
Audio-SAG (Figure~\ref{SAG_variants}(a)) (simulating Melodist~\cite{melodist24acl})                       & 47s  & 0.2216 & 0.1850 & 0.3064 & 2.9052 & 2.8348 & 2.6605 & 2.6613 & 2.6195 \\
MIDI-SAG 47s detected MIDI+Beat (Figure~\ref{SAG_variants}(b))    & 47s  & 0.6342 & 0.5550 & 0.2453 & 3.3032 & 3.1942 & 3.0425 & 3.0998 & 2.9353 \\
MIDI-SAG 47s (Figure~\ref{SAG_variants}(c))                      & 47s   & 0.9534 & 0.7700 & 0.2387 & 3.5473 & 3.4055 & 3.2921 & 3.3306 & 3.1434 \\
MIDI-SAG 90---120s (Figure~\ref{SAG_variants}(d))             & 90--120s     & \textbf{0.9134} & \textbf{0.8200} & 0.2820 & \textbf{3.6480} & \textbf{3.3957} & \textbf{3.4355} & \textbf{3.4341} & \textbf{3.2500} \\
\bottomrule
\end{tabular}
}
\end{table*}

Since the realized vocal pitch may naturally deviate from the quantized MIDI score, in our implementation, we extracted the pitch condition directly from the SVS output using the vocal pitch estimation model RMVPE~\citep{wei2023rmvpe}. In contrast, the harmonization model derives a chord progression from the groundtruth vocal MIDI.

We employed a more extensive set of time-varying controls than the original MuseControlLite. 
As shown in Table~\ref{tab:conditioning-signals}, we addressed the discrepancy between training (where ground-truth accompaniment is available) and inference (where signals must be derived from vocals audio or symbolic melody). 
Our preliminary experiments (cf. Table~\ref{tab:ablation}) confirmed that excluding chord or rhythm conditions led to unstable harmony and poor temporal alignment. Specifically, we resolve the failure of standard beat trackers on dry vocals by deriving rhythm directly from the vocal MIDI at inference time. To ensure structural coherence, all conditioning boundaries are aligned with downbeats or section boundaries.

To generate the full backing track, we employed a non-linear sequential approach to bypass the 47s generation limit. We first synthesized the first verse section without audio conditioning, then generated the intro using the verse as a ``backward reference'' (cf. Section \ref{completeness}). Subsequent segments were generated section-by-section, each conditioned on the previously synthesized audio window, until the outro. Finally, the completed accompaniment was mixed with the singing voice via summation and peak normalization.


\section{Experiment 1: Score-to-Song Generation}\label{experiment1}

Experiment 1 evaluates the score-to-song capability of all the SAG settings illustrated in Figure~\ref{SAG_variants}, using the same SVS module. Evaluation is conducted on a 200-sample test set generated with GPT-5, where each prompt specifies lyrics (verse, chorus, and outro) together with stylistic metadata. We first use CSL-L2M~\cite{chai2025csl} to generate vocal MIDI from the lyrics, and then synthesize the corresponding singing voice using our SVS system.
To assess the coherene of the generated backing track accross non-vocal and vocal sections, we append ``two bars of silence'' at the beginning to serve as the \emph{intro} part for the vocal track. For performance evaluation, we focus on rhythmic conistency, tonal/harmonic consistency, and aesthetics.

Following \citet{wu2024music}, we adopt \textbf{Rhythm F1} to evaluate alignment between the detected beats of the vocal and backing track, using All-in-One~\citep{kim2023all} for beat detection. A beat from the backing track is considered correct if it falls within 70\,ms of any beats from the vocal track. We also measure \textbf{Key Accuracy}, where a prediction is regarded as correct only when both the pitch class and mode (major/minor) match the reference. These metrics allow us to isolate the contribution of symbolic MIDI controls to the structural and harmonic stability of the generated outputs. In addition, we use \textbf{SongEval}~\cite{yao2025songevalbenchmarkdatasetsong} for aesthetic evaluation. 
SongEval scores an input song from 1 to 5 along five dimensions: Coherence, Musicality, Memorability, Clarity, and Naturalness.

The results are shown in Table~\ref{tab:midi_sag_results}.
For the first baseline, we evaluate TTM (SAO~\cite{evans2025stable}) without vocal input and additional training. This setting shows that omitting vocal information leads to rhythm and key misalignment, despite achieving the highest CLAP score~\cite{wu2024largescalecontrastivelanguageaudiopretraining}. We further fine-tune TTM with vocal audio, denoted as \textit{Audio-SAG}, to simulate the approach adopted in Melodist~\cite{melodist24acl}. Although Audio-SAG yields a slight improvement in key accuracy, rhythm alignment remains poor.

When vocal MIDI, vocal beat, and vocal pitch contour are extracted from vocal audio~\cite{openvpi_some, heydari2022singing}, and chord progression is provided by Accomontage2~\cite{yi2022accomontage2} based on the detected vocal MIDI, Rhythm F1, Key Accuracy, and SongEval scores all improve, at the cost of a slight decrease in CLAP~\cite{wu2024largescalecontrastivelanguageaudiopretraining}. Furthermore, when the vocal audio is synthesized from a given vocal MIDI, the vocal beat can be obtained directly from the MIDI file, and the ground-truth vocal MIDI enables higher-quality chord progression estimation. Under this setting, all evaluation metrics further improve except CLAP~\cite{wu2024largescalecontrastivelanguageaudiopretraining}.

Since MIDI-SAG demonstrates promising results for 47-second generation in a single pass, we further evaluate it on longer songs of 90--120 seconds using the audio continuation strategy described in Section~\ref{implement}. The last row of Table~\ref{tab:midi_sag_results} indicates that our method remains stable and consistent under this long-form audio continuation.



\begin{table}[t]
\centering
\small
\setlength{\tabcolsep}{4pt} 
\caption{Result of Experiment 2, comparing our MIDI-SAG and existing SAG baselines. For fairness, we use only vocal audio input (i.e., without ground-truth vocal MIDI) here.}
\label{tab:sag_comparison}
\begin{tabular}{lccc}
\toprule
 & AnyAccomp & FastSAG & \makecell[c]{Our MIDI-SAG\\w/o ground truth\\vocal MIDI} \\
\midrule
Training data (hrs)        & 8k         & 3k         & 2.5k       \\
Base generation length            & 10s        & 10s        & \textbf{47s}        \\
Longer generation           & $\times$  & $\times$  & $\checkmark$ \\
Text control capability          & $\times$ & $\times$  & $\checkmark$ \\
\midrule
APA~\citep{ grachten2025accompaniment} $\uparrow$                    & 0.457      & 0.000      & \textbf{0.595}       \\
\bottomrule
\end{tabular}
\end{table}

\begin{table*}[t]
  \centering
  \caption{Objective evaluation results of Experiment 3 for lyrics-to-song generation. $\uparrow$/$\downarrow$ indicates the higher/lower the better.}
  \label{obj}
  \setlength{\tabcolsep}{4pt}
  \resizebox{\textwidth}{!}{
  \begin{tabular}{l r r cc cccc ccccc}
    \toprule
    \multirow{2}{*}{\textbf{Model}} &
    \multirow{2}{*}{\makecell{\textbf{Total}\\\textbf{Parameters}}} &
    \multirow{2}{*}{\makecell{\textbf{Training}\\\textbf{Data (hrs)}}} &
    \multicolumn{2}{c}{\textbf{}} &
    \multicolumn{4}{c}{\textbf{Audiobox}$\uparrow$} &
    \multicolumn{5}{c}{\textbf{SongEval} $\uparrow$} \\
    \cmidrule(lr){4-5}\cmidrule(lr){6-9}\cmidrule(lr){10-14}
     &  & 
     & \textbf{CLAP $\uparrow$} & \textbf{PER $\downarrow$}
     & \textbf{CE} & \textbf{CU} & \textbf{PC} & \textbf{PQ}
     & \textbf{Coherence} & \textbf{Musicality} & \textbf{Memorability} & \textbf{Clarity} & \textbf{Naturalness} \\
    \midrule
    Suno v4.5
      & --- & ---
      & \textbf{0.417} & 0.290
      & 7.339 & 7.766 & 5.333 & 8.036
      & \textbf{4.198} & \textbf{4.011} & \textbf{4.174} & \textbf{4.034} & \textbf{3.939} \\
    ACE-Step
      & 3.5B & 100K
      & 0.368 & 0.238
      & 7.209 & 7.642 & 5.820 & 7.948
      & 3.449 & 3.214 & 3.203 & 3.216 & 3.162 \\
    DiffRhythm
      & 1.3B & 60K
      & 0.323 & 0.325
      & 7.530 & \textbf{7.791} & \textbf{6.336} & 8.189
      & 3.740 & 3.419 & 3.595 & 3.512 & 3.354 \\
    LeVo
      & 2B & 110K
      & 0.229 & 0.617
      & 7.565 & 7.674 & 4.993 & \textbf{8.295}
      & 3.392 & 3.272 & 3.198 & 3.265 & 3.155 \\
    \textbf{Ours}
      & 1.3B & 2.5K
      & 0.282 & \textbf{0.213}
      & 7.590 & 7.712 & 6.294 & 8.240
      & 3.648 & 3.396 & 3.436 & 3.434 & 3.250 \\
    \bottomrule
  \end{tabular}
  }
\end{table*}



\section{Experiment 2: MIDI-SAG without Groundtruth Vocal MIDI}


Experiment 2 evaluates the fundamental accompaniment generation capability of MIDI-SAG against existing audio-SAG baselines. Specifically, we benchmark our MIDI-SAG system (without ground-truth MIDI) against existing audio-SAG baselines on the standard \textbf{10-second accompaniment generation} task, where vocal MIDI is unavailable. We restrict the comparison to 10-second segments with continuous vocals to ensure fairness with existing models, which do not natively support longer durations or intermittent vocal silence.
We select FastSAG~\citep{chen2024fastsag} and AnyAccomp~\citep{zhang2025anyaccomp} as baselines, while excluding SingSong~\citep{donahue2023singsong}, VersBand~\citep{zhang2025versatile}, Melodist~\cite{melodist24acl}, and MelodyLM~\cite{li2024accompanied} because their code or checkpoints are not publicly available. For our method, we use the setting described in Figure~\ref{SAG_variants}(c), where vocal MIDI~\cite{openvpi_some} and beat information~\cite{heydari2022singing} are extracted automatically from the vocal audio. As the evaluation metric, we adopt Accompaniment Prompt Adherence (\textbf{APA})~\citep{grachten2025accompaniment}, where higher scores indicate better accompaniment quality and alignment. We use the MUSDB18~\citep{rafii2017musdb18} test set for evaluation and its training set as the reference set. For fairness, all audio samples are segmented into 10-second chunks.

 As shown in Table~\ref{tab:sag_comparison}, our model achieves the highest APA score among all evaluated methods, demonstrating superior alignment with the vocal track, despite being trained on a smaller dataset, highlighting the efficiency of symbolic conditioning. In contrast, FastSAG~\citep{chen2024fastsag} fails to generate meaningful accompaniment, likely due to its sensitivity to source separation artifacts in the training vocals. These results confirm that integrating symbolic controls through MuseControlLite~\cite{tsai2025musecontrollite} adapters effectively adapts the SAO backbone to the SAG task.

\section{Experiment 3: Lyrics-to-Song Generation}

Experiment 3 evaluates the bonus task of lyrics-to-song generation for producing complete songs of 90--120 seconds. We compare our compositional pipeline against representative end-to-end baselines, including Suno v4.5~\cite{suno_v4_5_2025}, ACE-Step~\cite{gong2025ace}, DiffRhythm~\cite{ning2025diffrhythm}, and LeVo~\cite{lei2025levo}. We use the same evaluation set as in Experiment~1, where the lyrics and metadata are generated by GPT-5, in order to reduce the chance that evaluation songs overlap with the training data of the compared systems. Since different baselines require different input formats, we reformat the metadata for each model while preserving the same semantic content.

Table~\ref{obj} presents the objective evaluation results. Among the open-source systems, our method achieves the lowest phoneme error rate (PER), and also outperforms Suno on this metric, indicating that the proposed compositional pipeline produces more intelligible and temporally aligned singing vocals. We attribute this advantage to the use of a dedicated SVS module, which explicitly models lyric-to-vocal rendering instead of jointly generating vocals and accompaniment in a single end-to-end stage. In contrast, our CLAP score is lower than those of Suno, ACE-Step, and DiffRhythm, suggesting that although our system is strong at vocal rendering, it still lags behind large-scale end-to-end systems in overall text--audio semantic alignment.

The Audiobox metrics show a more nuanced picture. Our method achieves the best CE score, the second-best PC score, and the second-best PQ score, while remaining close to the leading systems on CU. These results suggest that our model is competitive in objective audio quality, even though it is trained with substantially less data than the strongest baselines. For SongEval, Suno remains the strongest model across all five dimensions. Compared with the open-source baselines, our method ranks behind DiffRhythm on coherence, musicality, memorability, clarity, and naturalness, but remains consistently stronger than ACE-Step and LeVo on most of these metrics. Overall, the objective results indicate that our approach is particularly effective at lyric intelligibility and reasonably competitive in perceptual quality, while still leaving room for improvement in holistic prompt-conditioned song generation.

Table~\ref{subjective} reports the subjective listening results. Consistent with the objective evaluation, Suno is preferred by listeners on all five dimensions, revealing a clear gap between the commercial closed-source system and the open-source alternatives. Among the open-source models, our system ranks above DiffRhythm and LeVo in \textit{Overall Preference}, although it remains below ACE-Step. For \textit{Lyrics Adherence}, our model achieves the highest score among all open-source systems, slightly surpassing ACE-Step. This finding is broadly consistent with the strong PER result in Table~\ref{obj}, and suggests that listeners perceive the proposed pipeline as producing clearer and better-aligned vocal delivery.

A notable strength of our system is \textit{Voice Naturalness}, where it achieves the highest subjective score among the open-source methods. This result further supports the benefit of using a dedicated SVS component in the compositional pipeline. On the other hand, our model is weaker in \textit{Musicality} and \textit{Structure Clarity}, where it trails Suno, ACE-Step, and DiffRhythm. This suggests that while modular decomposition helps vocal quality and lyric rendering, global musical flow and large-scale arrangement coherence remain challenging for the current system.

Taken together, the results in Tables~\ref{obj} and~\ref{subjective} reveal a clear trade-off. Compared with large-scale end-to-end systems, our approach does not yet maximize overall aesthetic preference or text--audio alignment; however, it offers strong lyric intelligibility and voice naturalness with a far more data-efficient and controllable compositional design. These findings support the promise of adopting a compositional pipeline integrating MIDI-SAG as a practical framework for structured song generation, while also highlighting future directions for improving long-range musicality, structural clarity, and style-level controllability.

\begin{table}[t]
  \centering
  \caption{Subjective evaluation results of Experiment 3 for lyrics-to-song  generation (mean\,$\pm$\,confidence interval); higher values indicate better performance. Bold denotes the highest value, and underlining indicates the second highest.}
  \label{subjective}
  \setlength{\tabcolsep}{4pt} 
  \resizebox{\linewidth}{!}{
  \begin{tabular}{l ccccc}
    \toprule
    \textbf{Model} &
    \thead{\textbf{Overall}\\\textbf{Preference}} &
    \thead{\textbf{Lyrics}\\\textbf{Adherence}} &
    \textbf{Musicality} &
    \thead{\textbf{Voice}\\\textbf{Naturalness}} &
    \thead{\textbf{Structure}\\\textbf{Clarity}} \\
    \midrule
    Suno v4.5      & \textbf{4.297$\pm$0.194} & \textbf{4.375$\pm$0.188} & \textbf{4.344$\pm$0.201} & \textbf{4.188$\pm$0.195} & \textbf{4.297$\pm$0.184} \\
    ACE-Step       & \underline{2.953$\pm$0.249} & 3.656$\pm$0.244 & \underline{3.172$\pm$0.271} & 2.672$\pm$0.280 & \underline{3.266$\pm$0.252} \\
    DiffRhythm     & 2.469$\pm$0.231 & 2.922$\pm$0.192 & 2.859$\pm$0.235 & 2.703$\pm$0.238 & 2.688$\pm$0.239 \\
    LeVo           & 2.297$\pm$0.275 & 2.188$\pm$0.239 & 2.641$\pm$0.256 & 2.484$\pm$0.262 & 2.531$\pm$0.235 \\
    \textbf{Ours}  & 2.656$\pm$0.287 & \underline{3.672$\pm$0.243} & 2.703$\pm$0.242 & \underline{3.109$\pm$0.270} & 2.516$\pm$0.262 \\
    \bottomrule
  \end{tabular}%
  }
\end{table}

\begin{table}[t]
\centering
\caption{Ablation study on conditions for MIDI-SAG for score-to-song generation.}
\label{tab:ablation}
\small
\begin{tabular}{@{}lccc@{}}
\toprule
\textbf{Setting} & \textbf{Chord F1} & \textbf{Key Acc} & \textbf{Rhythm F1} \\
\midrule
w/ all conditions            & 0.9006 &0.79 &0.8339 \\
w/o chord        & 0.3908 & 0.21 & 0.7870  \\
w/o rhythm       & 0.8914 & 0.79  & 0.4279     \\
w/o structure    & 0.8957 & 0.79 & 0.8317     \\
w/o audio        & \textbf{0.9027} & \textbf{0.84}  & \textbf{0.8442}     \\
w/o vocal pitch contour & 0.5930 & 0.69  & 0.4319     \\
\bottomrule
\end{tabular}
\end{table}

\section{Ablation Study on Conditioning Signals}
Table \ref{tab:ablation} 
presents an ablation study clarifying the contribution of each conditioning signal to MIDI-SAG controllability. We observe that the scores are highest when the model is conditioned only on non-audio inputs. In this setting, the model is freed from reconciling with ambiguous reference audio, allowing it to strictly follow the remaining time-varying conditions. However, this simplification can introduce acoustic abruptness detectable by listeners, whereas the reference audio helps smooth transitions. 
Removing the chord condition greatly lowers both Chord F1 and Key Accuracy. This suggests the model derives the key implicitly from the chord progression. Conversely, excluding the vocal pitch contour leads to a performance drop across all metrics, indicating that SAG relies heavily on this condition for temporal and harmonic grounding.

\section{Discussion}
While our experiments demonstrate the effectiveness of the proposed compositional pipeline, the inherent risk of error propagation across sequential stages warrants discussion. 
We mitigate this risk through targeted implementation strategies at each modular interface.
To ensure vocal quality, we employ a register-check mechanism that validates the generated melody against the specialized vocal range of the SVS model prior to synthesis. Furthermore, the MIDI-SAG module is trained on conditions extracted from diverse real-world audio recordings.
Because these extracted signals naturally contain varying levels of noise, the model is forced to develop a \emph{robust interpretation} of its inputs during the learning process.
Consequently, our system exhibits significant resilience to minor inaccuracies from preceding stages, as the model’s training effectively prepares it to handle imperfect conditioning signals.

While current generations target 90--120s structural forms, this is not a fundamental constraint but a deliberate focus to evaluate structural coherence within a low-resource training setting. Although our system currently trails large-scale models, our modular foundation can be further enhanced by scaling data and parameters in future work, offering a transparent and editable path for long-form song generation.

\begin{table}[t]
\centering
\caption{Comparison of song generation systems. The length limit shown for our method is due to the limitation of CSL-L2M~\cite{chai2025csl}. In contrast, under singing accompaniment generation or score-to-song settings, our method is able to generate longer backing tracks.}
\label{tab:system_comparison}
\small
\setlength{\tabcolsep}{3pt}
\renewcommand{\arraystretch}{1.15}
\begin{tabular}{lcccr}
\toprule
\textbf{Model} & \shortstack{\textbf{Vocal MIDI}\\\textbf{generation}} & \shortstack{\textbf{Score-to-}\\\textbf{song}} & \shortstack{\textbf{Lyrics-to-}\\\textbf{song}} & \shortstack{\textbf{Max}\\\textbf{length}} \\
\midrule
Melodist~\cite{melodist24acl}        & $\times$ & $\checkmark$ & $\times$ & 10s \\
MelodyLM~\cite{li2024accompanied}    & $\checkmark$ & $\checkmark$ & $\checkmark$ & 12s \\
VersBand~\cite{zhang2025versatile}   & $\checkmark$ & $\checkmark$ & $\checkmark$ & 12s \\
Ours                                 & $\checkmark$ & $\checkmark$ & $\checkmark$ & 90--120s \\
\midrule
Diffrhythm~\cite{ning2025diffrhythm} & $\times$ & $\times$ & $\checkmark$ & 285s \\
ACE-Step~\cite{gong2025ace}           & $\times$ & $\times$ & $\checkmark$ & 280s \\
Levo~\cite{lei2025levo}              & $\times$ & $\times$ & $\checkmark$ & 270s \\
\bottomrule
\end{tabular}
\end{table}

\section{Conclusion}

This paper introduces a compositional song pipeline centered on MIDI-informed singing accompaniment generation (MIDI-SAG), bridging the gap between automated music synthesis and professional creative control. By transitioning from an end-to-end approach to a modular framework, we demonstrate that symbolic musical priors, such as vocal MIDI and explicit chord roadmaps, are essential for maintaining rhythmic and harmonic coherence in long-form compositions. Our approach specifically addresses the ``long-form bottleneck,'' enabling the generation of structurally complete songs that include instrumental intros, bridges, and outros, which are often unachievable by conventional audio-only models. As summarized in Table~7, our method occupies a unique position among existing systems by supporting vocal MIDI generation, score-to-music, and lyrics-to-music within a single compositional framework, while extending the generation length of prior modular score-aware approaches~\cite{melodist24acl,li2024accompanied,zhang2025versatile} from around 10--12\,s to 90--120\,s in the current song generation setting.

While our system is an early exploration of compositional long-form song generation, the results confirm that high-fidelity, temporally aligned accompaniment can be achieved through a data-efficient architecture trained on consumer-grade hardware. By leveraging specialized pre-trained modules, our framework remains flexible and future-proof, allowing for the integration of evolving SVS and accompaniment technologies. Ultimately, this work suggests that the 
future of music AI 
can lie in developing modular, controllable systems that respect the structural integrity of musical form and the creative authorship of the user.

\section{Acknowledgment}
This work is supported by grants from the Ministry of Education (MOE) of Taiwan (for Taiwan Centers of Excellence), the National Science and Technology Council of Taiwan (NSTC 114-2628-E-002-013-MY3), and Google Asia Pacific.  The authors are also grateful to Dr. Li Chai for her assistance on combining the lyrics-to-melody model (CSL-L2M) into the lyrics-to-song pipeline.

\appendix

\section{Appendix}
\label{appendix:implementation}

\subsection{Lyrics-to-Melody Generation}
\label{appendix:lyrics2melody}

We employ the CSL-L2M model~\cite{chai2025csl} to map Chinese lyrics to vocal MIDI scores. The architecture is a Transformer decoder utilizing an in-attention mechanism~\cite{wu2023musemorphose} for fine-grained control. Generation is conditioned on global attributes (key and emotion), sentence-level structure tags, and twelve statistical musical attributes (e.g., pitch variance, note density, and syllable-to-note alignment) that strengthen lyric--melody coupling. While users can specify emotion and key manually, the statistical attributes are derived from a reference MIDI track. The model achieves optimal performance when the reference track and target lyrics share similar section structures and word counts.

To facilitate generation without user-provided MIDI, we curated a reference bank of 1,000 attribute sets. For a given set of input lyrics, we select the most compatible candidate using a weighted penalty score
\[
P = 0.4\,P_{\text{sent}} + 0.4\,P_{\text{prof}} + 0.2\,P_{\text{struct}},
\]
where lower scores indicate higher compatibility. Here, \(P_{\text{sent}}\) penalizes discrepancies in total line count (optionally rejecting candidates with fewer lines than the target); \(P_{\text{prof}}\) is the mean absolute difference between per-line token counts---treating each visible Chinese character as one token---after padding the shorter sequence with its median and scaling by the maximum observed token count; and \(P_{\text{struct}}\) compares section tags mapped to integers, counting position-wise mismatches and adding a penalty for extra sections, normalized by the longer sequence length.

\subsection{Melody Harmonization}
\label{appendix:midi2chord}

Since chord progressions serve as primary time-varying controls, the inference pipeline must supply compatible harmonic sequences. We harmonize the vocal MIDI scores produced by CSL-L2M using AccoMontage2~\citep{yi2022accomontage2}. To ensure a natural musical flow, we prepend a 4-bar instrumental intro; the chord sequence for this intro is generated by duplicating the chords from the first four bars of the melody.

This harmonization stage provides the essential chordal grounding that allows the singing-accompaniment generator to produce coherent backing tracks. To maintain user agency, the system supports manual overrides: if the automatically generated progression is unsatisfactory, users may provide their own chord sequences or perform partial edits on the generated results.

\subsubsection{Implementation Details on Singing Voice Synthesis}
\label{appendix:svs}

We integrate a MIDI-conditioning embedding to align each phoneme with its corresponding pitch and duration in the MIDI score. To ensure natural vocal quality, we perform a register-matching optimization before synthesis. Given the vocal MIDI track, we evaluate potential octave shifts $\Delta \in \{-12,0,+12\}$ against both male (lower) and female (higher) vocal profiles. The optimal configuration $(\text{singer}, \Delta)$ is selected by maximizing the number of notes within the profile’s comfortable tessitura while simultaneously minimizing the magnitude of the octave shift $|\Delta|$. This ensures the generated vocals remain within a realistic performance range while preserving the melodic intent of the original MIDI as closely as possible.

To reconstruct the final waveform audio, we fine-tuned a Parallel WaveGAN vocoder~\citep{yamamoto2020parallel} specifically on our singing voice dataset. The training was conducted for one week on a single NVIDIA RTX 3090 GPU, optimizing the model to capture the nuances and high-frequency details characteristic of vocal performances.

\begin{figure*}[t]
    \centering
    \includegraphics[width=\textwidth]{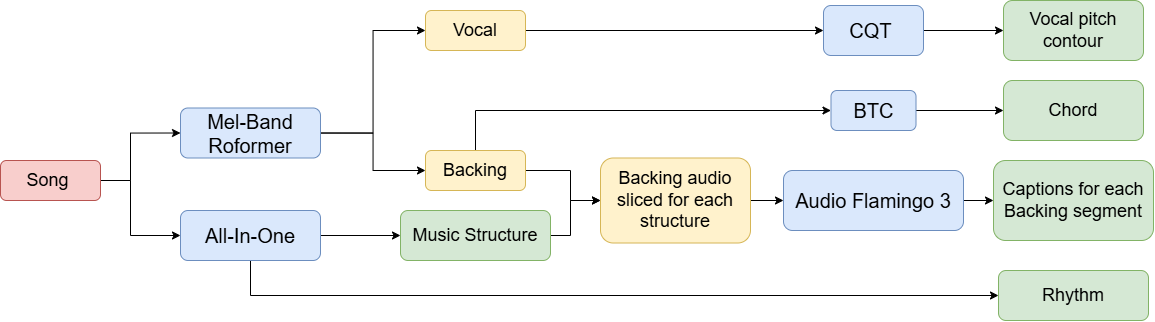}
    \caption{The data-preprocessing pipeline to curate data for fine-tuning Stable Audio Open to implement our MIDI-informed singing accompaniment generation (MIDI-SAG) model.}
    \label{fig:data-preprocessing}
\end{figure*}

\subsection{Singing Accompaniment Generation}
\label{appendix:sag}

To fine-tune the Stable Audio Open backbone, we curated a specialized dataset of Mandarin pop music following the pipeline illustrated in Figure~\ref{fig:data-preprocessing}. The preparation involves source separation to isolate vocal and backing tracks~\citep{wang2023mel} and automated audio captioning~\citep{goel2025audio}. To provide the model with granular guidance, we extract a comprehensive set of time-varying conditioning signals, including chords~\citep{ParkCJKP19}, local key~\citep{SchreiberM19_CNNKeyTempo_SMC}, rhythmic features and structural tags~\citep{kim2023all}, and vocal pitch contours~\citep{hou2025editing}. The source audio was retrieved from public web sources and is maintained exclusively as an internal dataset for academic research purposes, yet we commit to share the dataset metadata. We fine-tuned the model on NVIDIA RTX 3090 GPUs using an effective batch size of 108, training for a total of 9 days to ensure convergence across the diverse conditioning signals.

\subsection{Conditioning Signals}
\label{appendix:conditions}

We utilize an extensive set of controls, building upon the framework of MuseControlLite~\citep{tsai2025musecontrollite}, to address the specific challenges of low-resource, long-form song generation. A key distinction in our approach lies in the source of these signals: during training, conditioning signals are extracted directly from the ground-truth accompaniment to ensure high precision. Conversely, at inference time, these conditions must be derived solely from the synthesized vocal audio or the symbolic vocal MIDI. This transition requires the model to be resilient to the slight variations inherent in predicted signals.

\textbf{Vocal pitch contour.}
To capture melodic nuance, we first isolate the vocal track using Mel-Band RoFormer~\citep{wang2023mel}. We then extract prominent pitch information using the top-4 Constant-Q Transform (CQT) method proposed by \citet{hou2025editing}. During the training phase, this contour is derived from the ground-truth isolated vocals to provide a precise melodic anchor. At inference, the signal is extracted directly from the SVS-generated singing, allowing the accompaniment module to track the synthesized vocal performance with high fidelity.

\textbf{Rhythm.}
Our pilot study shows that existing beat tracking models do not work well on isolated vocal audio. For example, BeatNet~\citep{heydari2021beatnet} achieved a Rhythm F1 score of only 0.3449 in our evaluation. Consequently, we adopt a dual-strategy approach. During training, we use All-In-One~\citep{kim2023all} to extract beat and downbeat timestamps from the ground-truth backing tracks. These are converted into binary indicator sequences of shape \((T, 1)\), where \(T\) represents the number of time frames. We then apply a Gaussian filter to these sequences to produce smooth rhythm activation curves. At inference time, as audio-based tracking remains unreliable for singing voices, we derive beat and downbeat timings directly from the quantized MIDI generated by the CSL-L2M model~\citep{chai2025csl}, which outputs melodies in 4/4 time.

\textbf{Chord.}
Our preliminary experiments indicated that the SAG model generates unstable harmony and weak progressions when deprived of explicit chordal conditioning. To ensure harmonic stability, we implement a chromagram-based approach. During training, we apply a chord detector~\citep{ParkCJKP19} to the isolated backing tracks and encode the detected progressions as 12-bin chromagrams, representing pitch-class membership over time. At inference, these harmonic cues are supplied by the AccoMontage2~\citep{yi2022accomontage2} harmonization module.

\textbf{Structure.}
To organize the song’s narrative and energy flow, we extract section labels and timestamps using All-In-One~\citep{kim2023all}. We retain a standardized set of labels—including intro, verse, chorus, bridge, solo, break, inst, and outro—while discarding truncated fragments at the audio boundaries. Furthermore, we utilize AudioFlamingo3~\citep{goel2025audio} to generate section-level text captions that provide high-level semantic guidance.

\begin{figure*}[t]
    \centering
    \includegraphics[width=\textwidth]{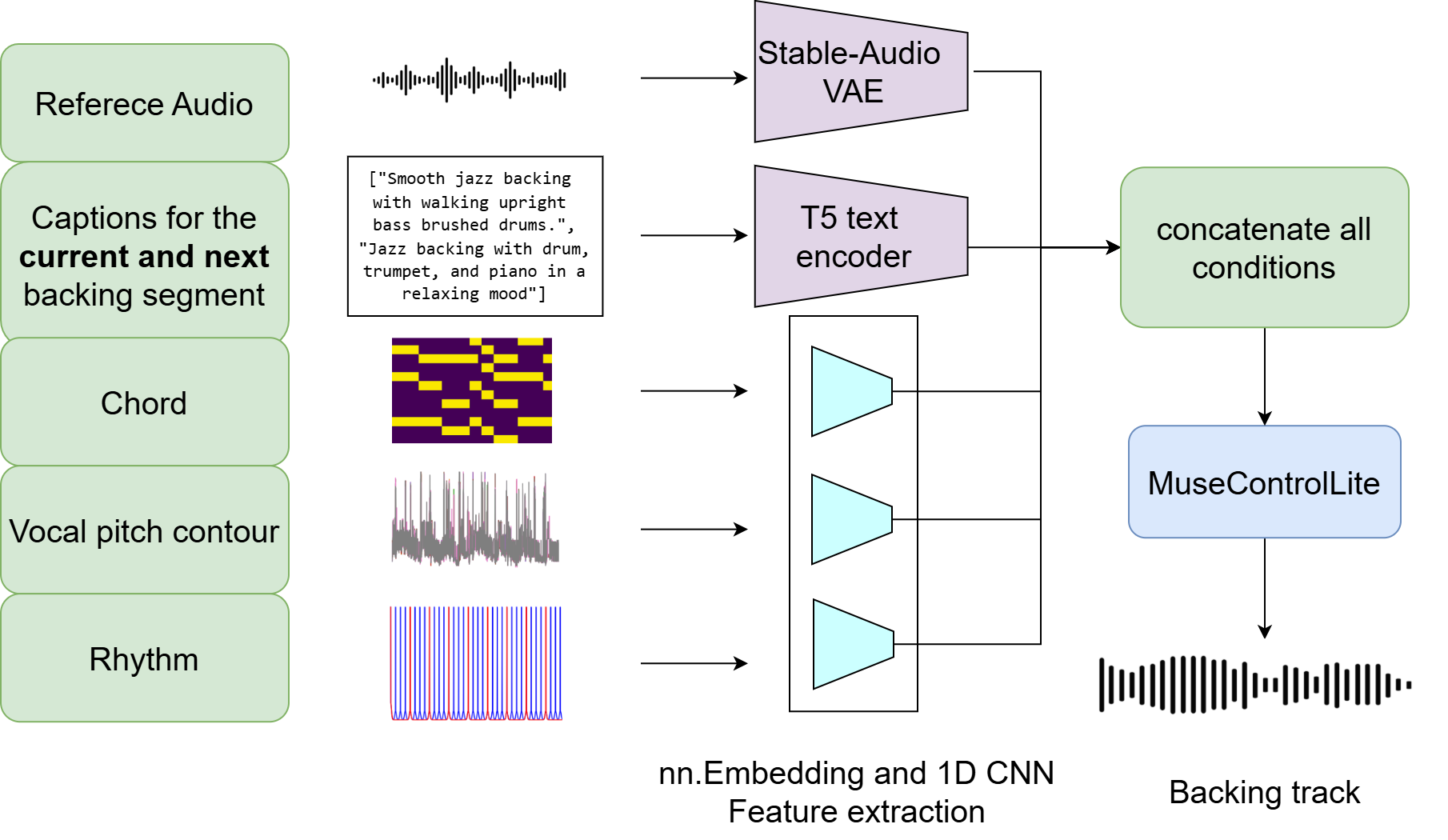}
    \caption{Augmenting Stable Audio Open for singing accompaniment and audio continuation. The architecture utilizes the MuseControlLite framework to integrate multi-modal conditioning signals, enabling precise singing-accompaniment alignment and seamless long-form audio continuation.}
    \label{fig:learning-time-varying}
\end{figure*}

\textbf{Reference audio.}
To facilitate structural completeness, we employ reference audio conditioning similar to MuseControlLite~\citep{tsai2025musecontrollite}. This allows the model to utilize inpainting and outpainting techniques for the generation of instrumental sections, ensuring stylistic consistency between synthesized and retrieved audio.

\subsection{Objective Evaluation Metrics for Song Generation}
\label{appendix:metrics}

\textbf{Lyrics Alignment.}
We employed Whisper ASR~\citep{radford2022robustspeechrecognitionlargescale} to transcribe the generated vocals and compared the transcriptions with the ground-truth lyrics. Alignment quality is measured using the phoneme error rate (PER). We first convert both the predicted and reference texts into their phoneme representations. Then, PER is computed as $\mathrm{PER} = \frac{S + D + I}{N}$, where $S$, $D$, and $I$ denote substitutions, deletions, and insertions, respectively. A lower PER indicates better alignment.

To verify the suitability of Whisper as an evaluation proxy for singing voice, we benchmarked its performance on the \textsc{CPOP} dataset~\citep{mirex2018alignment}, which consists of high-quality human vocal recordings. By comparing Whisper's transcriptions against the ground-truth lyrics, we obtained a PER of 0.059.

\textbf{Style Alignment.}
We utilized CLAP~\citep{wu2024largescalecontrastivelanguageaudiopretraining} to compute the cosine similarity between audio embeddings of the generated music and embeddings of the text prompts.

\textbf{Aesthetics Evaluation.}
We used Audiobox-Aesthetics~\citep{tjandra2025metaaudioboxaestheticsunified} to provide an automatic aesthetics assessment of the generated songs.

\textbf{SongEval.}
We use SongEval~\citep{yao2025songevalbenchmarkdatasetsong} to measure structural clarity, memorability, musical coherence, and overall musicality.

\textbf{Controllability.}
To evaluate whether MIDI-SAG successfully aligns with the given conditions, we extract rhythm, chord, and key features from the generated backing audio using the same procedure as in training. Following \citet{wu2024music}, we use the F1 score to evaluate rhythm alignment.

\begin{table}[t]
  \centering
  \caption{Inference latency per module for a 120-second song.}
  \label{tab:inference-time}
  \begin{tabular}{l r}
    \toprule
    \textbf{Module} & \textbf{Time (s)} \\
    \midrule
    Lyrics-to-melody & 10 \\
    Melody harmonization & 0.2 \\
    Singing voice synthesis & 3 \\
    Singing accompaniment generation & 40 \\
    \bottomrule
  \end{tabular}
\end{table}

\begin{table*}[t]
  \centering
  \caption{Ablation study demonstrating the effectiveness of unfreezing self-attention layers for audio continuation.}
  \label{tab:musecontrol_ablation}
  \footnotesize
  \setlength{\tabcolsep}{4pt}
  \renewcommand{\arraystretch}{1.0}
  \begin{tabular}{lcccc}
    \toprule
    Method & FD$\downarrow$ & KL$\downarrow$ & CLAP$\uparrow$ & Smoothness$\uparrow$ \\
    \midrule
    MuseControlLite 
    & 111.58 & 0.2160 & 0.3622 & \textbf{-0.2734} \\
    
    MuseControlLite w/ unfrozen self-attention layers
    & \textbf{109.62} & \textbf{0.1794} & \textbf{0.3961} & -0.3529 \\
    \bottomrule
  \end{tabular}
\end{table*}

\subsection{Additional Experimental Results}

To evaluate the computational efficiency of our pipeline, we measured the inference latency for each component on a single NVIDIA RTX 3090 GPU. The results, summarized in Table~\ref{tab:inference-time}, detail the time required for each stage—from initial melody generation to final singing-accompaniment synthesis—for a standard full-length song.

The results in Table~\ref{tab:musecontrol_ablation} demonstrate that partially unfreezing the self-attention layers of the SAO backbone significantly enhances generation quality.

We also tested the vocal beat detector~\cite{heydari2022singing} on a 250-hour test set separated from the training set, and the beat F1 score was 0.73.

\begin{table*}[t]
\centering
\footnotesize
\setlength{\tabcolsep}{4.5pt}
\renewcommand{\arraystretch}{1.15}
\newcolumntype{L}[1]{>{\raggedright\arraybackslash}p{#1}}
\newcolumntype{Y}{>{\raggedright\arraybackslash}X}

\begin{tabularx}{\linewidth}{L{0.7cm} L{2.6cm} L{2.9cm} Y Y}
\toprule
\textbf{Stage} &
\textbf{Component} &
\textbf{Base model / source} &
\textbf{Upstream training (prior work)} &
\textbf{Our additional training (this work)} \\
\midrule
\textbf{1} &
Lyrics $\rightarrow$ Vocal MIDI &
CSL-L2M~\cite{chai2025csl} &
300 hours lyrics--melody pairs; trained with 1$\times$V100 for 24 hours &
None \\
\addlinespace[2pt]

\textbf{2} &
SVS (MIDI $\rightarrow$ vocals) &
FastSpeech~\cite{ren2020fastspeech} &
None &
10 h licensed internal (2 singers); GPU: 1$\times$3090 for 8 days; trained from scratch \\
\addlinespace[2pt]

\textbf{3} &
Backing generator backbone &
Stable Audio Open~\cite{evans2025stable} &
7.3k hours licensed data; 36224 A100 GPU hours (VAE + Diffusion Transformer) &
None \\
\addlinespace[2pt]

\textbf{3b} &
MIDI-SAG conditioning module (adapter) &
MuseControlLite &
None &
2.5k h Mandarin pop; GPU: 1$\times$3090 for 9 days \\
\addlinespace[2pt]

\textbf{4} &
Harmony / chord extraction or harmonization &
AccoMontage2~\cite{yi2022accomontage2} &
None; template matching and dynamic programming &
None \\
\addlinespace[2pt]

\textbf{5} &
Beat / tempo tracking (for silent spans) &
\citet{heydari2022singing} &
None &
2.5k h Mandarin pop; GPU: 1$\times$3090 for 12 hours \\
\bottomrule
\end{tabularx}

\caption{Component-level provenance and training effort. We explicitly separate upstream pretraining from incremental training performed in this paper.}
\label{tab:provenance_training_effort}
\end{table*}

\subsection{Our Training Data Distribution}

Our training data primarily consists of Mandarin pop music, as our goal was to generate music from Chinese vocals. The genre distribution is based on structure-level text prompts predicted by AudioFlamingo~\cite{goel2025audio}.

\subsection{Regarding Training Resources Used in the \emph{Compositional} Pipeline}

Table~\ref{tab:provenance_training_effort} shows the pretraining effort and the incremental training cost of each module. The total amount of training data used across all components is 10k hours.

\subsection{Input Examples of Our Compositional Song Pipeline}

\begin{figure*}[t]
   \centering
    \includegraphics[width=.65\textwidth]{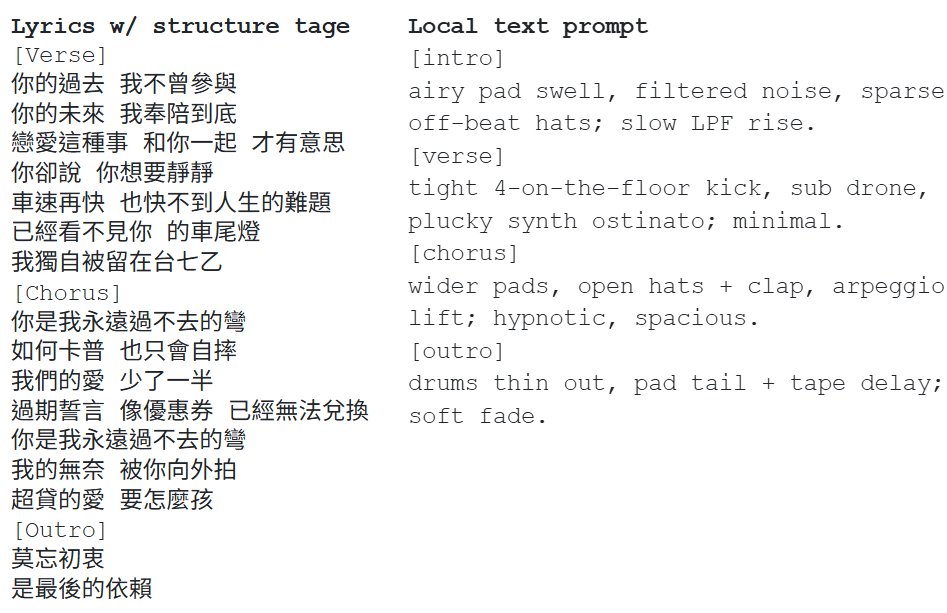}
    \caption{The input lyrics are the same as those used by other end-to-end song generation models, but our method supports Mandarin only due to the constraint of CSL-L2M~\citep{chai2025csl}. The text control of our method could be either a single global style prompt or different local style prompts for different segments.}
    \label{fig:lyrics_demo}
\end{figure*}

The input of our compositional song pipeline is shown in Figure~\ref{fig:lyrics_demo}. It is basically the same as other end-to-end song generation models, but our model supports using different text prompts to control different musical structures.

\FloatBarrier

\bibliographystyle{ACM-Reference-Format}
\bibliography{sample-base}

\end{document}